\documentstyle[preprint,aps]{revtex}
\begin{document}
\preprint{DRAFT hep-th/9406202}

\title{Quantization of the Gravitational Constant in
Odd-Dimensional Gravity}
\addtocounter{footnote}{1}

\author {J. Zanelli
\thanks{Electronic address: cecsphy@lascar.puc.cl}}
\address{ Centro de Estudios Cient\'{\i}ficos de Santiago,
Casilla 16443, Santiago 9, Chile\\
and Departamento de F\'{\i}sica, Facultad de Ciencias,
Universidad de Chile, Casilla 653, Santiago, Chile}

\date{June 1994}
\maketitle

\begin{abstract}

It is pointed out that the action recently proposed by Ba\~nados
et al. for gravitation in odd dimensions higher (and lower) than
four, provides a natural quantization for the gravitational
constant. These theories possess no dimensionful parameters and
hence they may be power counting renormalizable.

\end{abstract}

Gravitation in dimensions greater than two is best described by
the so-called Lovelock action \cite{lovelock}. This is a linear
combination of the dimensional continuations to $D$ dimensions
of all the Euler classes of dimensions $2p<D$ \cite{zumino,tz}.
The Lovelock lagrangian could be defined by the previous
statement, but it can also be derived in three other seemingly
independent ways: i) It is the most general invariant
constructed out of the metric and curvature that yields second
order covariant field equations \cite{lovelock}; ii) It is the
most general local $D$-form invariant under tangent space
rotations, constructed out of the vielbein, the spin connection
and their exterior derivatives without using the Hodge-* dual
\cite{regge}; iii) It is the most general low energy effective
theory of gravity that can be derived from string theory
\cite{zwiebach}.

In a recent article, Ba\~nados et al.\cite{banados}, have
considered a particular form of the Lovelock action which
results from embedding the group of tangent space rotations,
$SO(D)$, into $SO(D+1)$. \footnote{For brevity here $SO(D)$ will
denote the group of rotations in $D$ dimensions or any of its
complex extensions $SO(p,q)$, with $p+q=D$. We shall make the
distinction below when commenting on the spacetime signature and
the Wick rotation.} For odd dimensions there is a particular
choice that makes the action invariant under $SO(D+1)$, whereas
even $D$ that possibility does not exist.

The proposed Lagrangian for $D=2n-1$ dimensions reads

\begin{equation}
L_{2n-1}=\kappa\sum_{p=0}^{n-1} \frac{1}{D-2p} \left( \begin{array}{c}
n-1 \\
p \end{array} \right) l^{2p-D} \epsilon_{a_1 \cdots a_{2n-1}}
R^{a_1 a_2}\wedge \cdots \wedge R^{a_{2p-1} a_{2p}}
\wedge e^{a_{2p+1}}\wedge \cdots \wedge e^{a_{2n-1}}.
\label{0}
\end{equation}

Here $\kappa$ is the gravitational constant analogous to
Newton's in $D=4$, $l$ is a constant with dimensions of length,
$R^{ab}$ is the curvature two form and $e^a$ are the vielbein.
As it is shown in Ref. \cite{banados}, this is the
Euler-Chern-Simons density. That is, $L_{2n-1}$ is a
($2n-1$)-form whose exterior derivative is the Euler class
${\cal E}_{2n}$ for ($2n$)-dimensional manifolds,

\begin{equation}
d\wedge L_{2n-1} = \kappa {\cal E}_{2n},
\label{1}
\end{equation}
with
\begin{equation}
{\cal E}_{2n}= \epsilon_{A_1 \cdots A_{2n}} \bar{R}^{A_1 A_2}\wedge
\cdots \wedge \bar{R}^{A_{2n-1} A_{2n}}
\label{2}
\end{equation}
where $A, B, C,...= 1, ..., 2n$. Here $\bar{R}^{AB}$ is the
$2n$-dimensional curvature 2-form associated to the (anti-)
de Sitter group,

\begin{equation}
\bar{R}^A_B = \left[\begin{array}{cc}
               R^a_b -\epsilon l^{-2} e^a_b  &  \epsilon l^{-1} T^a \\
        -l^{-1} T_b  &    0
\end{array} \right],
\label{2.3}
\end{equation}
where $T^a =de^a +\omega^a_b \wedge e^b$ is the torsion 2-form.

$L_{2n-1}$ is the analog of the Pontryagin Chern-Simons form
encountered in gauge theories. In gravity there is also a
Pontryagin form --often called the Hirzebruch class-- in $D=4n$
and a corresponding Chern-Simons density in $D=4n-1$ (see e.g.,
\cite{mardones}), but we will not consider them here
as they cannot be dimensionally continued.

For a nonbelian gauge theory in 2+1 dimensions with gauge group
$G$ the existence of large gauge transformations within a non
trivial homotopy class implies the quantization of the coupling
constant, $g$, that multiplies the Chern-Simons action
\cite{jackiw,percacci}. Roughly speaking, if $\pi_3(G) \neq 0$, $g$
must be quantized.

The same argument applied to asymptotically flat gravitation
theory in 2+1 dimensions does not lead to the quantization of
Newton's constant because in that case the relevant group is
$ISO(2,1)$, and $\pi_3(ISO(2,1))=0$ \cite{percacci}.

On the other hand, the theories considered in \cite{banados} have
a cosmological constant, their solutions are not asymptotically
flat and hence the relevant groups are anti-de Sitter
$SO(2n-2,2)$, so the non quantization argument does not apply
here.

Our argument for the quantization of $\kappa$ however, does not
rely on the existence of lage gauge transformations but on the
possibility of rewriting the $(2n-1)$-dimensional gravity
action as a topological theory on a $2n$-dimensional manifold
whose boundary is the spacetime one wants to describe. This is
similar to the standard discussion leading to the quantization
of the coupling constant in the WZW theory \cite{Witten}.

Let us consider the particular case of a compact
($2n-1$)--dimensional, simply connected manifold $M$ that is the
boundary of some $2n$--dimensional orientable manifold $\Omega$
[For definiteness, $M$ could be taken to be $S^{2n-1}$]. Then,
by Stokes' theorem, the action for $M$ can be expressed as

\begin{equation}
S_{\Omega}[M] = \kappa \int_{\Omega} {\cal E}_{2n} .
\label{3}
\end{equation}

Obviously, there is a large freedom in the choice of $\Omega$, as
there are infinitely many ways to extend $M$. However, since the
action $S_{\Omega}[M]$ is to describe the dynamical properties
of $M$, it is reasonable to demand that the observables of the
system should be insensitive to changing in eq.(\ref{3})
$\Omega$ by a different $2n$-manifold, $\Omega'$, with the same
boundary ($M$) \cite{milnor},

\begin{equation}
\partial \Omega =M= \partial \Omega'.
\label{4}
\end{equation}
Also,

\begin{eqnarray*}
S_{\Omega}[M] &=& \kappa \int_{\Omega} {\cal E} =
             \kappa(\int_{\Omega} {\cal E} - \int_{\Omega'} {\cal E}) +
             \kappa\int_{\Omega'} {\cal E}. \\
        &=&\kappa(\int_{\Omega} {\cal E} + \int_{-\Omega'} {\cal E}) +
S_{\Omega'}[M].
\end{eqnarray*}

The first term on the right hand side of the last equality is $\kappa$
times the Euler class of the manifold formed by joining $\Omega$ and
$-\Omega'$ smoothly along $M$. The minus sign accounts for the fact that
the orientation of one of the two halves must be reversed in order for
their union to possess a well defined orientation throughout. Thus, we
finally have

\begin{equation}
S_{\Omega}[M] = \kappa \chi[\Omega \cup -\Omega'] + S_{\Omega'}[M].
\label{5}
\end{equation}

Although  the action of any classical system is defined modulo
an arbitrary additive constant, quantum mechanically this
constant must be an integer multiple of Planck's constant $h$ so
that the path integral of the system be unaffected. This
implies, in particular, that under continuous transformations of
the fields the additive constant cannot change.

On the other hand, the replacement $\Omega \rightarrow \Omega'$
could not be attainable through a continuous transformation if
$\Omega$ and $\Omega'$ have different topologies. This means
that the difference $S_{\Omega'}[M] - S_{\Omega}[M]$ must be an
integer multiple of $h$, and one concludes from(\ref{5}) that
since $\chi$ is an integer, $\kappa$ must be quantized. [The
same argument doesn't hold for gravity in $2n$ dimensions
because there is no analog of eq. (\ref{1}) in that case and
hence the even dimensional action cannot be written as the
integral of an exact $2n+1$-form.]

Our point rests on the assumption that there exist a manifolds
of the form $\Omega \cup -\Omega'$ with a nonzero Euler
characteristic. It is actually not difficult to envissage many
examples of this type, e.g., $\Omega$ and $\Omega'$ can be two
halves of a $2n$-sphere with any number of "handles" attached to
each hemisphere. A different question is whether $\Omega \cup
-\Omega'$ could be a classical solution for a $2n$-dimensional
theory. This question is related to the existence of instantons
like the D'Auria-Regge solution \cite{dauria} in four
dimensional Euclidean gravity (that instanton, however, is not a
solution of pure gravity but requires torsion and matter
fields). The existence of such solutions is not required for the
validity of our argument and does not concern us here since we
are taking the point of view that the fundamental theory is the
one defined in $2n-1$ dimensions.

The quantization argument is valid regardless of the signature
of spacetime. This is because the Euler characteristic is a
topological invariant and hence insensitive to changes in the
signature of the metric. The action is constructed entirely out
of form fields which are independent of the coordinates and
therefore invariant under coordinate changes, including Wick
rotations.  Thus, both in the hyperbolic and the Euclidean
signature the integrand of the path integral is $exp(\frac{-i}
{\hbar}S[M])$, where $S[M]$ is the (real) action constructed as
in (\ref{0}).

It might seem paradoxical that the action be insensitive to the
signature of the metric, especially in view of the explicit
dependence of $L_{2n-1}$ on the vielbein. The paradox is
resolved by noting that the Wick rotation must be performed
simultaneously on the spacetime coordinates and on the tangent
space, as this is the only consistent way to maintain the
relationship between the spacetime metric and that of its
tangent space (soldering), $g_{\mu \nu} =\eta_{ab} e^a_{\mu}
e^b_{\nu}$. Since in the end all (spacetime and tangent space)
indices are contracted in the Lagrangian, no extra factors of
$i$ appear in the Euclidean action.

Actions whose construction require the Hodge *-dual (like
$\partial_{\mu}\phi \partial^{\mu}\phi \sqrt{|g|} d^4x$ or
$F^{\mu \nu}F_{\mu \nu}\sqrt{|g|} d^4x$) do change under Wick
rotations because they explicitly involve the epsilon symbol,
which is a pseudoscalar and hence transforms with an additional
factor of $i$. The usual Euclidean action for gravity, $I_E=i
\int d^4x_E \sqrt{g_E} R$, carry an extra $i$ that can be
viewed, in the language of forms, as resulting from the
substitution $\epsilon_{a b c d} \rightarrow i\epsilon_{a b c
d}$ when going from $SO(3,1)$ to $SO(4)$. Our point of view here
is that this is not necessary (in odd dimensional spacetimes), but
if one insists in introducing an $i$ when the group is changed,
the definition of the theory should be such that the Euclidean
sector gives an imaginary phase for the path integral. This is
consistent with the requirement that time reversal be equivalent
to conjugation in the path integral (see e.g., \cite{barnatan}).

The proposal of Ref. \cite{banados} for a gravitation theory in
odd dimensions possesses a number of interesting features and,
as we have shown here, its coupling constant is quantized under
standard assumptions. An interesting consequence of the
quantization of the gravitational constant is that the Hilbert
space for 2+1 gravity with cosmological constant has
finite-dimensional unitary representations \cite{hayashi}.

In addition, the action (\ref{0}) written in terms of the
rescaled vielbein $e^a \rightarrow l e^a$, has no dimensionful
constants and all fields have canonical dimension one.
Furthermore, the fields $\omega^a_b$ and $e^a$ are different
components of the connenction and the action describes a bona
fide $SO(D+1)$ gauge system. The corresponding quantum theory
would be renormalizable by power counting and possibly finite.
Witten has shown this to be the case in three dimensions using
the fact that the diffeomorphism constraints can be solved
classically for $D=3$, leaving only a discrete set degrees of
freedom to be quantized \cite{witten}.

For $D>3$ however, the gravitational field propagates and the
construction of a quantum theory might be a formidable task. In
fact, a construction analogous to that of Ref. \cite{witten}
probably doesn't exist at all. Recently, Ba\~nados and Garay
have developed a first order Hamiltonian formalism for
(\ref{0}). In that form the theory possesses a consistent
constraint algebra, which suggests that a quantum version of the
system might eventually be constructed \cite{b-g}.

\vspace{2cm}


Many useful discussions with E. Gozzi, M.  Henneaux, K. S.
Narain, R. Percacci, C. Teitelboim and G. Thompson are
warmly acknowledged. The author is particularly grateful to
M. Ba\~{n}ados for many enlightening comments and to S. Carlip
for correspondence that clarified several subtle points. This work
was supported in part by Grants Nos.  293.0007/93 and
193.0910/93 of FONDECYT (Chile), by a European Communities
research contract, and by institutional support to the Centro de
Estudios Cient\'{\i}ficos de Santiago provided by SAREC (Sweden)
and a group of chilean private companies (COPEC, CMPC, ENERSIS).
The author wishes to express his gratitude to Professor Abdus
Salam, the International Atomic Energy Agency and UNESCO for
hospitality at the International Centre for Theoretical Physics,
Trieste.


\end{document}